\documentstyle{mn}
\newcommand{\ltsima} {$\; \buildrel < \over \sim \;$}
\newcommand{\gtsima} {$\; \buildrel > \over \sim \;$}
\newcommand{\lta} {\lower.5ex\hbox{\ltsima}}
\newcommand{\gta} {\lower.5ex\hbox{\gtsima}}

\def\a{\alpha}
\def\b{\beta}
\def\g{\gamma}
\def\G{\Gamma}
\def\d{\delta}

\def\l{\lambda}
\def\t{\theta}
\def\m{\mu}
\def\n{\nu}

\def\s{\sigma}
\def\refitem{\par\parskip 0pt\noindent\hangindent 20pt}

\title[The kinetic power of GRS 1915+105]
{The bulk kinetic power of the jets of GRS 1915+105}

\author[M. Gliozzi, G. Bodo \& G. Ghisellini]
{M. Gliozzi$^1$, G. Bodo$^2$ \& G. Ghisellini$^3$\\
$^1$ Dipartimento di Fisica Generale di Torino, Via P.Giuria, I-10125 
Torino, Italy\\
$^2$ Osservatorio Astronomico di Torino, Strada Oservatorio 20, I-10025 
Pino Torinese, Italy\\
$^3$ Osservatorio Astronomico di Brera, V. Bianchi, 46, I-23807 Merate,
Italy}

\begin{document}
\maketitle

\begin{abstract}
We calculate the minimum value of the power in kinetic bulk motion
of the galactic superluminal source GRS 1915+105.
This value far exceeds the Eddington luminosity for accretion onto a 
black hole of 10 solar masses.
This large value severely limits the possible carriers of the kinetic 
luminosity at the base of the jet, and favours a jet production and 
acceleration controlled by a magnetic field whose value, at the base 
of the jet, exceeds $10^8$ Gauss.
The Blandford and Znajek process can be responsible of the 
extraction of the rotational energy of a Kerr black hole,
if lasting long enough to provide the required kinetic energy.
This time, of the order of a day, implies that the process must operate
in a stationary, not impulsive, mode.
\end{abstract}


\section{Introduction}
GRS 1915+105 was discovered in 1992 with the WATCH telescope on board the
GRANAT satellite (Castro-Tirado et al. 1992). 
A radio counterpart was subsequently identified and bipolar outflows 
with apparent superluminal motions were observed; the standard 
interpretation of this phenomenon in terms of relativistic jets 
(Rees 1966) places the source at a distance $D$=12.5 kpc 
at an angle $i=70^\circ$ to the line of sight (Mirabel \& Rodriguez 1994). 
The source cannot be observed in the optical band, due to the heavy 
extinction in the Galactic plane, but 
the X-ray luminosity often well above the Eddington luminosity for a
neutron star suggests the presence of a $\sim$10 solar mass black hole.
Other circumstantial evidence comes with 
the similarities with the other galactic superluminal
source GRO J1655--40, which has been shown unambiguously to harbor a
compact object of $\simeq 7 M_{\odot}$ (Orosz \& Bailyn 1997).
Since its discovery, GRS 1915+105 has displayed an extraordinary richness in 
variability in the X-ray band (Greiner, Morgan \& Remillard 1996, Chen, Swank
\& Taam 1997, Belloni et al. 1997), in the IR and radio bands (Fender et al. 
1997, Pooley \& Fender 1997). 
Recent (Eikenberry et al. 1997, Mirabel et al. 1997) simultaneous
multiwavelength observations of this superluminal source show a strict 
link between activity in the inner accretion disk and plasma ejections.

In this paper we estimate the minimum kinetic power associated with
the ejections and investigate the possible energy transport 
mechanism from the compact source to distances where the radio blobs
are resolved, showing that the more suitable solution implies the 
presence of a strong magnetic field at the footpoint of the jet. 

In section 2 we summarize the existing radio and IR data of the ejection
events. In section 3 we directly calculate the minimum kinetic power
during the ejection.
In section 4 we investigate the possible origin of the kinetic power.
In section 5 we summarize the main results of the paper and discuss their 
consequences.

\section{Observational constraints during the outflows}
Radio observations show that GRS 1915+105 in its active state is
characterized by the increase of the flux level from $S_{\n}\le 10$ mJy
to $S_{\n}\sim 100$ mJy, the so called {\it plateau} state, on the top
of which radio {\it flares} are superimposed, with durations from days
to a month and fluxes up to a few Jy (Foster et al. 1996). The spectra 
during the plateau state are generally flat or inverted,
$\a\le 0$ (with $S_{\n}\propto\n^{-\a}$), probably indicating
synchrotron self-absorption. 
On the other hand the spectra of radio flares reveal a transition 
from optically thick to thin ($\a> 0$) emission during 
the rise stage, with spectra harder ($\a\sim 0.5$)
when the fluxes reach the maximum and softer ($\a\sim 1$) during the
subsequent decay, generally interpreted as indicator of synchrotron
radiation of an expanding radio cloud.

Observations with the Very Large Array and multiwavelength campaigns 
(with VLA and UKIRT simultaneously) pointed out that some
strong radio flares are related with the ejection of radio clouds.
There are two different kind of ejections: 
1) {\it major ejections} as the prominent radio outburst of March 1994 
(Mirabel \& Rodriguez 1994), where the ejecta actually moved for several 
weeks along a direction forming an angle $\t=70^\circ$ to the line of sight, 
with bulk velocity of $0.92 c$ and expansion at $\sim 0.2 c$, with a
spectral index change from $\a=0.49$ (when the blobs could not be resolved) 
to $\a=0.84$ (when the two condensations had moved apart); and 
2) so-called {\it "baby jets"} (Eikenberry et al. 1997, Mirabel et al. 1997): 
quasi-periodic oscillations in the radio flux, coupled with similar
oscillations in X-ray and IR bands and interpreted as scaled-down
(in space and time) ejections of synchrotron emitting expanding blobs.
The analysis of QPOs in IR and radio bands clearly shows that
there is a time shift of flux peaks, with short wavelengths peaking
first: the IR (2 $\mu$m) peak precedes the 2 cm peak, that precedes 
the 3.6 and 6 cm peaks.

It must be emphasized that in one occasion (July 1995)
also in the near-infrared K band 
($\l=2.2\m$m) a jet was observed, with the same position angle of the 
radio jet, and near-infrared magnitude of $K$=13.9 (slightly brighter than
the total source magnitude at its weakest measured value of $K$=14.3) and
separated from the central source by 0.3'' (Sams et al. 1996).

\section{Minimum kinetic power condition}
In order to deduce the minimum kinetic luminosity $L_{k}$ related to
a major ejection we consider the March 1994 event 
for which the observational data are the most exhaustive. 
Since we are interested more in power rather than in energy estimates, 
we follow an alternative way with respect to the standard one, 
that consists to determine first the internal energy of the blob via 
the minimum energy criterium and then to obtain an estimate of the kinetic 
power, by making some (highly uncertain) assumptions about the 
energization time.

We instead directly calculate the kinetic power, equal
to the energy flux through a cross section of the jet
(see also Ghisellini \& Celotti, 1998).
This energy flux can be carried by particles and by the toroidal 
magnetic field, and correspondingly we have
$$
L_{k,p}=\pi R^2 \Gamma^2 \b c n^\prime 
<\g> m_e c^2 \left(1+{m_+\over m_e<\g>}\right)
\eqno(1a)
$$
$$
L_{k,B}=\pi R^2\G^2\b cU_B
\eqno(1b)
$$

\noindent
where $m_+$ is either $m_p$ in the case of a ``normal" $e-p$ plasma or 
$<\g>m_{e^+}$
for $e^\pm$ pairs, $\G$ is the Lorentz factor of the bulk motion,
$R$ is the cross section radius of the jet, $n'$
in the comoving particle density, $U_B$ is the magnetic energy density 
measured in the comoving frame and $<\g>$ is the mean Lorentz factor
of the electrons, as measured in the comoving frame.
A lower limit on $n^\prime$ can be estimated by the observed synchrotron
emission $L_{syn}$ of the blob. 
Assuming a spherical emitting volume of radius $R$,
the number density of leptons producing the observed radiation is
$$
n^\prime \, = \, {9 L_{syn}\over 2 <\g^2>\d^4\s_T cB^2R^3}
\eqno (2)
$$
where $<\g^2>$ is averaged over the relativistic electron distribution,
and $\d=[\G(1-\b\cos\theta)]^{-1}$ is the Doppler factor.
The $B^{-2}$ dependence of the estimated particle density allows
to minimize the total power $L_{k,p}+L_{k,B}$
with respect to the magnetic field,
since both the bulk Lorentz factor $\Gamma$ (=2.55) 
and the viewing angle $\theta$ (=72$^\circ$) are known:
$$
{\partial\over\partial B} \left( L_{k,p}+L_{k,B}\right) \equiv 0
\eqno (3)
$$
yielding the value of the magnetic field $B_{min}$ corresponding to
the minimum power:
$$
B_{min}=\left[{36\pi L_{syn} f m_e c^2\over R^3\d^4\s_T c} \right]^{1/4},
\eqno(4)
$$
where $f$ is a parameter directly related to the shape of the electron
energy distribution, and therefore to the shape
of the synchrotron radiation they emit:
$$
f\, \equiv \, {<\g> \over <\gamma^2>}\, 
\left(1+ {m_+\over m_e <\g>}\right).
\eqno(5)
$$
Since the radio emission is a power law with spectral index $\alpha\sim0.5$,
the particle distribution $N(\gamma)\propto \g^{-2}$ between $\gamma_1$ 
and $\gamma_2$, yielding 
$f\sim \ln(\gamma_2/\gamma_1)/\gamma_2
\{1+m_+/m_e[\gamma_1\ln(\gamma_2/\gamma_1)]\}$.
With $R\sim 7\times 10^{15}$ cm (geometric average of the size of 
blob of GRS 1915+105, assuming a distance of 12.5 kpc), 
$L_{syn}=10^{33}$  erg s$^{-1}$ and $\b=0.92$ (corresponding to $\G=2.55$), 
we obtain $B_{min,ep}=0.12$ G and $B_{min,e^{\pm}}=0.036$ G
if $f=1.8$ ($\gamma_1=1$ and $\gamma_2=10^3$, as required for production
of the observed synchrotron photons at $\nu\sim 300$ GHz).
A low energy cut-off in the particle distribution would decrease
somewhat $B_{min}$ ($B_{min,ep}=0.054$ G and $B_{min,e^{\pm}}=0.031$ G
for $\gamma_1=30$), while the high energy cut-off is consistent 
with the production of the observed radiation at 300 GHz (with $B = B_{min}$).
However, the dependence of $B_{min}$ on the extremes of the electron 
distribution is rather weak ($B_{min}\propto f^{1/4}$).

Note that the value $B_{min,ep}=0.12$ is almost a factor 3 greater than the 
one obtained by Mirabel \& Rodriguez (1995) and Liang \& Li (1995),
while it substantially agrees with the value estimated from the 
requirement of optical transparency of the plasmoid with respect 
to synchrotron self absorption (Atoyan \& Aharonian 1997).

With a magnetic field value equal to $B_{min}$, the particle kinetic power 
and
the Poynting flux are nearly equal, and the total power  
$$
L_{k,tot}\, =\, {3\Gamma^2\b c \over 2\delta^2}
\left[{\pi m_ec^2 R L_{syn} f  \over  \sigma_T c} \right]^{1/2}
\eqno(6)
$$
is, respectively for $e-p$ plasma and $e^{\pm}$ pairs,
$$
\, \sim \, 
3.3\times 10^{40}\, {\rm erg \,s^{-1}\qquad and \qquad} 
\sim \,2.9\times 10^{39}\, {\rm erg \,s^{-1}}
$$ 
This is the minimum kinetic luminosity involved in major ejection events,
calculated assuming $\gamma_1=1$ and $\gamma_2=10^3$
($L_{k,tot,ep}\sim 6.4\times 10^{39}$ erg s$^{-1}$ 
and $L_{k,tot,e^{\pm}}\sim 2.1\times 10^{39}$ erg s$^{-1}$ if $\gamma_1=30$).

Setting $B=B_{min}$ in equation (2) 
the total numbers of emitting particles in the magnetic cloud are respectively
$N_{e^-}^\prime =5.9\times 10^{47}$ and $N^{\prime}_{e^\pm} =6.9\times 10^{48}$,
corresponding to internal energies
$$
E_i\, \equiv \, m_ec^2\int \g N(\g)d\g\, =N<\g>m_ec^2\
$$
$$
\sim\, 3.3\times 10^{42}\,{\rm erg},\qquad \qquad
\sim\, 3.9\times 10^{43}\,{\rm erg}
\eqno(7)
$$
These internal energies can be compared with the corresponding kinetic energies 
in bulk motion $E_{k}$ given by
$$
E_{k}=L_{k}\, t\, \simeq \, L_{k}\,{R\over \b c}\,
$$
$$
\sim 8.5\times 10^{45}\, {\rm erg} \qquad {\rm and} \qquad\sim 7.4\times 10^{44}
{\rm erg}
\eqno(8)
$$
where $t$ is the time needed for the blob to cross the jet section.

\section{Origin of the kinetic luminosity}
The kinetic power calculated above refers to the radio emitting blob, 
at a huge distance from the putative black hole and accretion disk.
The most economic assumption is that this power is approximately
conserved along the jet, with very small dissipation giving rise, e.g., to
the random particle energy responsible for the emission.
If not, we would have to assume that at the base of the jet an even larger
value of $L_{k}$ is produced.
In principle there are several possible energy carriers: $e^\pm$ pairs, 
$p-e^-$, pure magnetic field or a mixture of these components.
Consider also that the particles can in principle be ``hot" or ``cold":
at large distances at least some of the electrons are ``hot'' 
(indeed, in our estimates we assumed that all the electrons are ``hot'',
since this is the most economic assumption), however these ``hot'' particles
cannot come directly from the inner region (see below), but need to be accelerated
or reaccelerated, therefore for the inner jet energy carriers one has to
consider both cases. 
Furthermore, we have to note that in the ``hot" case, the particle 
random energy, increasing the
mass, can contribute to the kinetic power.
Let us examine the possible cases.

\subsection{``Cold" $e^\pm$ pairs}
If the kinetic power is carried by a pure $e^\pm$ cold pair plasma, we
can calculate the corresponding pair density and scattering
optical depth at some jet radius $R$ close to the base of the jet 
$$
\tau_\pm\, =\, \sigma_T R n_\pm \, =\, 
{\sigma_T L_{k} \over \pi R \Gamma^2 \b m_ec^3 }
$$
$$
 \sim \, 4.2\times 10^2 \,{ L_{k}\over 2.9\times 10^{39}
{\rm erg~s^{-1}} } \, {10^7~ {\rm cm} \over R}
\eqno(9)
$$
since the annihilation time-scale is of the order of $R/(c\tau_\pm)$,
cold pairs cannot survive annihilation.

\subsection{``Hot" $e^\pm$ pairs}
If the pairs are relativistic, with average random Lorentz factor
$<\g>$, the above estimate of $\tau_\pm$ decreases by a factor $<\g>$.
In addition, the annihilation cross section decreases.
% However, the Comptonization $y$ parameter, defined as 
% $y=\max(\tau_\pm, \tau_\pm^2) <\gamma^2>$ increases, 
% and pairs would inevitably be cooled in the dense photon environment
% surrounding the inner jet.
However these relativistic pairs are embedded in a dense radiation
field, produced by the accretion disk, and they cool on time-scales
shorter than the dynamical time-scale.
In a region of size $R_d$ the radiation energy density 
due to the accretion disk luminosity $L_d$ produced within $R_d$
is of the order of $U_d=L_d/(4\pi R_d^2c)$.
The ratio between the inverse Compton cooling time and the dynamical time 
$R_d/c$ for a particle of Lorentz factor $\gamma$ is 
$$
{t_{IC} \over R_d/c} \, =\, {4\pi R_d m_ec^3 \over \sigma_T L_d \gamma}\,
\ll 1
\eqno(10)
$$
A large fraction of $L_{k}$ would be lost and converted into radiation,
contrary to what observed.
We therefore conclude that a pure $e^\pm$ pair plasma cannot carry 
the kinetic power, irrespective of whether it is hot or cold.

\subsection{ ``Normal" $e-p$ plasma}
The case of electrons with average random Lorenz factor $<\gamma>$
greater than $m_p/m_e$
and cold protons is analogous to the above case: inverse Compton
scattering of seed accretion disk photons cools the electrons in
a time shorter than $R_d/c$.

The case of ``hot", relativistic protons is instead immune to radiative
losses, and the main concern comes from the needed confining pressure.
If the kinetic power is carried by protons with an average Lorentz 
factor $<\gamma_p>$, the corresponding (comoving) pressure is of the
order of their energy density.
If the confinement is magnetic, the required value of the magnetic
field is
$$
B\, = \, \left( {8L_{k} \over R^2 \Gamma^2 \b c} \right)^{1/2}
\eqno(11)
$$
which is equal to the value that the magnetic field must have if
it is the main carrier of the kinetic power.

If both electrons and protons are cold, the electron scattering optical 
depth is a factor $m_p/m_e$ smaller than the one of equation (9), 
i.e. of the order of a few.
There are no severe limits in this case: the Compton
drag is not sufficient to slow down the plasma in the jet.
It is however useful to estimate the predicted power emitted by
the Compton drag process, because it can turn out to be important 
in sources where the jet is more aligned with the
line of sight.
For this simple estimate we will assume that the jet is Compton
thick (i.e. $\tau_T>1$), so that the effective cross section of the
process is the geometrical one, i.e. $\pi R^2$.
As before, we assume that the accretion disk produces the luminosity
$L_d$ within the region $R_d>R$, at the typical frequency $\nu_d$.
With these hypotheses, the observed bulk Compton luminosity $L_{bC}$ is
$$
L_{bC}\, \sim\, 
\left( {R\over R_d} \right)^2  {L_d \over 4} \Gamma^2 \delta^4
\eqno(12)
$$
Most of this luminosity is observed at the frequency 
$\nu \sim \delta\G\nu_d$.
With $R/R_d \sim 1/10$, 
$L_d \sim 10^{39}$ erg s$^{-1}$ and $\delta=0.57$
we obtain $L_{bC}\sim 1.7\times 10^{36}$ erg s$^{-1}$ and $\nu\sim \nu_d$.
This radiation is therefore unobservable in the case of the known
galactic superluminals, which are both observed with large viewing angles,
but may be very important for sources with $\delta\sim \Gamma \gg 1$
(see also Sikora et al., 1997).

\subsection{Poynting vector}
The value of the magnetic field needed to carry $L_{k}$ 
in the vicinity of the black hole is equal to equation (11),
if the bulk of the magnetic field is moving with $\G$,
while the $B$--value before acceleration is a factor $\G$ greater,
corresponding to $B\sim 3\times 10^8$ G at $R=10^7$ cm.

It must be remarked that the standard theory of accretion disks (Shakura
\& Sunyaev 1973) predicts that the maximum possible magnetic field
at a given $R/R_S$ (where $R_S$ is the Schwarzshild radius) in a 
radiation dominated disk is $B\propto (M/M_{\odot})^{-1/2}$, so that
microquasars will have magnetic fields $\sim 10^4$ stronger than those
in quasars. Therefore, the theoretical estimate $B\sim 10^2 - 10^4$ G 
obtained with the Blandford \& Payne model (Blandford \& Payne 1982)
near the massive black holes in active galactic nuclei, clearly 
show that in microquasars $B\sim 10^8$ G can easily be attained.

Note that this value of the magnetic field would be of the same magnitude 
of the magnetic field required by the Blandford--Znajek (1977) process to 
produce $L_{k}$ by extracting the rotational energy of a 10 solar mass 
Kerr black hole:
$$
L_{rot}\, =\, 10^{41} \, B_9^2 M_1^2\,\,\, {\rm erg~~ s^{-1}}
\eqno(13)
$$
where $B=10^9 B_9$ G and $M=10 M_1$ solar masses.

\vskip 0.5 true cm
From the above arguments we conclude that $e^\pm$ or relativistic
electrons with $<\g>$ greater than $m_p/m_e$ cannot carry $L_{k}$
in the inner region of the jet, while protons and the Poynting vector can.

An alternative solution could be that the energy, carried in the inner
region by ``normal" cold plasma or toroidal magnetic field, is converted
in $e^\pm$ pairs at large distance  
from the black hole and accretion disk.
Pairs are produced most efficiently through photon--photon collisions,
but this requires a powerful $\gamma$--ray continuum.
Since the maximum efficiency in converting pairs in this way is of the
order of 10 per cent (Svensson 1987), the high energy luminosity
has to exceed $10^{40}$ erg s$^{-1}$, to produce a kinetic power 
of the order of $10^{39}$ erg s$^{-1}$.
This luminosity is not observed.
 Therefore we conclude that $e^\pm$ pairs
do not play any role as energy carriers along the jet and the minimun
kinetic luminosity involved in major ejection events is given by the value 
estimated for $e-p$ plasma ($\sim 3\times 10^{40}~{\rm erg~ s^{-1}}$). 

There are no strong observational arguments to decide between protons and
the Poynting vector, unless we observe a new galactic superluminal at a small 
viewing angle.
Note however that if $L_{k}$ is initially carried by a large
magnetic field there is the possibility to tap a great reservoir
of energy (the rotational energy of the black hole) and 
to accelerate the plasma to relativistic speeds.

We therefore conclude that the kinetic power carried initially by the 
magnetic field is the most economic way to explain the observed energetics.
Differential accretion disk rotation or rotating black holes can amplify 
magnetic fields to the required values, and the magnetic field will 
then accelerate particles to relativistic speeds.

The amount of matter present at the base of the jet may not be negligible, 
especially if magnetic field lines help to channel particles from the disk 
to the jet.
We can compare the inflow rate $\dot M_{in}$ of the accretion
process with the outflow rate $\dot M_{out}$ necessary to account for the 
kinetic power, in particles, of the radio blob.
If we write the accretion luminosity, as usually, as 
$L_{acc}=\eta \dot M_{in} c^2$, and the kinetic power as 
$L_{k} = (\Gamma-1) \dot M_{out} c^2$, we derive

$$
\dot M_{out} \, =\, {\eta \over \Gamma -1} \, {L_{k} \over L_{acc} }\, 
\dot M_{in}
\eqno(14)
$$
Since $\eta\sim 0.1$ and $L_{k}\sim 10 L_{acc}$ we have that the
the inflowing and the outflowing mass rates are comparable.
This in turn suggests that most of the matter in the jet may come 
from the accretion disk.

\section{The ejection time}
Another important question concerns the duration of the ejection events.
When the blob becomes visible in the radio, it has a size of a few light days.
It seems unlikely that it corresponds to an ejection duration lasting
more than that, while a shorter ejection time may be possible.
In the latter case the requirement on the initial kinetic power 
correspondingly increases.
Note that direct radio observations of flare events do not directly 
constrain the ejection time, since the radio flux eventually produced 
in the first parts of the jet is heavily self absorbed.
Again, the minimum power requirements criterium favours ejection
events which last for $t_{out}\sim$ few days.
This corresponds to $\sim 10^9~R_{S,10}/c$, where $R_{S,10}$ is the 
Schwarzchild radius for a 10 solar mass black hole.
{\it The immediate consequence is that the ejection event 
must be considered as a continuous and stationary process.}
Scaling for a superluminal active galactic nucleus, we would
have, for a $10^9$ solar mass black hole, an ejection phase
lasting for $2\times 10^5$ years.

\section{Discussion}
We have obtained a reliable lower limit to the kinetic power corresponding 
to major ejection events in superluminal galactic sources,
in particular for GRS 1915+105. 
This limit is of the order of $3\times 10^{40}$ erg s$^{-1}$,
much greater than the observed radiative luminosity, which
is probably Eddington limited to values of the order of 10$^{39}$
erg s$^{-1}$, corresponding to a black hole of 10 solar masses.
This by itself suggests that the jet acceleration mechanism cannot be 
radiative.

We have investigated the role of $e^\pm$ pairs, of normal plasma
and of the magnetic field as energy carriers of the kinetic
power in the inner jet regions, excluding an important role
for the $e^\pm$ pairs, and favouring a scenario in which
the inital acceleration phase is controlled by a magnetic field
of the order of $10^8$ Gauss or more, able to channel and
accelerate accretion disk matter in the jet.

Conservation of kinetic power dictates that the injection time-scale 
is of the order of a day, a very long time-scale if measured in units 
of the light crossing time of a 10 solar mass Schwarzchild radius,
indicating a stationary, not impulsive, process.
Shorter injection times, albeit possible, correspond to larger kinetic 
powers, exacerbating the problem of how to obtain them.
The rough equality between the value of the magnetic field needed
to carry the kinetic power in the inner jet and the value necessary to tap 
the rotational energy of a Kerr black hole via the Blandford \& Znajek 
process can be regarded as circumstantial evidence that this process is 
indeed the one responsible for the jet formation and acceleration.
This process is candidate to power also the jets in radio--loud quasars,
and we can compare the results obtained here with the corresponding 
estimates of the kinetic power of AGN, as derived by Celotti, 
Padovani \& Ghisellini (1997) for a sample of radio--loud sources. 
For those AGNs, the above authors find a kinetic power between $10^{45}$ and 
$10^{48}$ erg s$^{-1}$, of the same order of the luminosity needed to ionize 
the broad line region of the same objects, and in agreement with the power 
required by the existence of the outer radio lobes.
In the AGN case the kinetic and the accretion luminosity are roughly equal, 
while in the galactic superluminal objects the kinetic power dominates.
Another, obvious, difference is the Lorentz factor of the bulk motion,
much greater in the AGN case.
This results in a ratio of the outflowing to infalling mass rate 
$\dot M_{out}/\dot M_{in}$ of order unity for the galactic superluminals 
and two orders of magnitude smaller for AGN.
These estimates makes the galactic superluminal sources the most efficient
engines to produce collimated relativistic bulk motion, with the possible
exception of gamma--ray bursts.

\section*{References}

\refitem Atoyan \& Aharonian 1997, MNRAS in press

\refitem Belloni, T., Mendez, M., King, A. R., van der Klis, M. \& van  
  Paradijs, J. 1997, ApJ 479, L145

\refitem Blandford, R.D \& Znajek, R.L. 1977, MNRAS, 179, 433

\refitem Blandford, R.D \& Payne, D.G. 1982, MNRAS, 199, 8833

\refitem Castro--Tirado, A.J., Brandt, S. \& Lund, S. 1992, IAU Circ. 5590

\refitem Celotti, A., Padovani, P. \& Ghisellini, G., 1997, MNRAS, 286, 415

\refitem Chen, X., Swank, J.H. \& Taam, R.E. 1997, ApJ 477, L41

\refitem Eikenberry, S.S., Mattheus, K., Morgan, E.H., Remillard, R.A. \&
  Nelson, R.W. 1997, ApJ in press

\refitem Fender, R.P., Pooley, G.G., Brocksopp, C. \& Newell, S.J. 1997
  MNRAS 290, L65

\refitem Foster, R.S., Waltman, E.B., Tavani, M., Har\-mon, B.A., Zhang, S.N.,
   Pa\-cie\-sas, W.S. \& Ghi\-go, F.D. 1996, ApJ 467, L81

\refitem Ghisellini G. \& Celotti A., 1998, submitted to MNRAS

\refitem Greiner, J., Morgan, E.H. \& Remillard, R.A. 1996, ApJ 473, L107

\refitem Landau, L.D. \& Lifshitz, E.M. 1963, Electrodynamics of Continuous Media,
   Pergamon, Oxford

\refitem Liang, E. \& Li, H. 1995, A\&A 298, L45

\refitem Meier, D. 1996, ApJ 459, 185

\refitem Mirabel, I.F. \& Rodriguez, L.F. 1994, Nature 371, 46

\refitem Mirabel, I.F. \& Rodriguez, L.F. 1995, 17th Texas Symp. on
  relativistic astroph. New York Acad. of Sc.

\refitem Mirabel, I.F., Dhawan, V., Chaty, S., Rodriguez, L.F., Marti, J.,
   Robinson, C.R., Swank, J.H. \& Geballe, T.R. 1997 A\&A in press

\refitem Orosz, J.A. \& Bailyn, C.D. 1997, ApJ 477, 876

\refitem Pooley, G.G., \& Fender, R.P. 1997, MNRAS in press

\refitem Rees, M.J. 1966, Nature 211, 468 

\refitem Sakura, N.I. \& Sunyaev, R.A. 1973 A\&A 24, 337

\refitem Svensson R., 1987, MNRAS, 227, 403

\refitem Sams, B.J., Eckart, A. \& Sunyaev, R.A. 1996, Nature 382, 47

\refitem Sikora, M., Madejski, G., Moderski, R. \& Poutanen, J., 1997, 
   ApJ 484, 108

\end{document}